# Energy Efficient Resource Allocation for Demand Intensive Applications in a VLC Based Fog Architecture

Wafaa B. M. Fadlelmula, Sanaa H. Mohamed*, *Member, IEEE*, Taisir E. H. El-Gorashi*, Jaafar M. H. Elmirghani, *Fellow, IEEE*

School of Electronic and Electrical Engineering, University of Leeds, Leeds, United Kingdom
* Department of Engineering, Faculty of Natural, Mathematical & Engineering Sciences, King's College London
elwbf@leeds.ac.uk, sanaa.mohamed@kcl.ac.uk, taisir.elgorashi@kcl.ac.uk, jaafar.elmirghani@kcl.ac.uk

**ABSTRACT**

In this paper, we propose an energy efficient passive optical network (PON) architecture for backhaul connectivity in indoor visible light communication (VLC) systems. The proposed network is used to support a fog computing architecture designed to allow users with processing demands to access dedicated fog nodes and idle processing resources in other user devices (UDs) within the same building. The fog resources within a building complement fog nodes at the access and metro networks and the central cloud data center. A mixed integer linear programming (MILP) model is developed to minimize the total power consumption associated with serving demands over the proposed architecture. A scenario that considers applications with intensive demands is examined to evaluate the energy efficiency of the proposed architecture. A comparison is conducted between allocating the demands in the fog nodes and serving the demands in the conventional cloud data center. Additionally, the proposed architecture is compared with an architecture based on state-of-art Spine-and-Leaf (SL) connectivity. Relative to the SL architecture and serving all the demands in the cloud, the adoption of the PON-based architecture achieves 84% and 86% reductions, respectively.

## 1. INTRODUCTION

Recently, we have witnessed an unprecedented number of devices being connected to the Internet. Based on Cisco Annual Internet Report (2018–2023) [1], 29.3 billion devices will be connected to the Internet by 2023. This increase will be associated with a demand for high date rates and timeliness exceeding the capabilities of the emerging 5G networks. The 6G networks vision promises increased data rates by further exploitation of the electromagnetic spectrum. Optical wireless frequencies offer a potential bandwidth exceeding 540 THz that can complement the Radio Frequency (RF) spectrum in access networks [2]. Several studies have proposed techniques to improve the achievable data rates of optical wireless communication (OWC) systems including adaptation of the beam power, angle, and delay [3], [4]. Visible light communication (VLC) is one of the promising OWC systems that uses light emitting diode (LEDs) or laser diodes (LDs) for indoor lighting and communication. For indoor applications, VLC provides high data rates of 25 Gbps and beyond [5]–[7] and enhanced security as light does not penetrate walls. VLC can also provide low-cost communication as it remains unregulated and unlicensed. Furthermore, VLC is an energy efficient technology as existing lighting infrastructure can be used for communication.

The exponential growth in traffic and processing demands is accompanied by an increase in power consumption. Network energy efficiency have been investigated extensively in the literature including proposing energy efficient architectures for data centers and core networks [8], [9], virtualization [10], [11], integration of renewable energy sources [12], [13], and content distribution optimization [14]. In the access network, passive optical networks (PONs) have proven their efficiency in reliably supporting high data rates at low power consumption. PONs have been proposed to provide backhaul connectivity in 5G network between the radio base stations and the network gateway [15], [16]. PONs were also proposed for data center interconnection (i.e., inter-rack communication and intra-rack communication) in [17]– [19]. Furthermore, PONs have the potential to improve the energy efficiency of fog computing [20] where 75% of the processing will be performed in fogs by 2025 [21].

In this paper, we investigate the use of PONs to provide backhaul connectivity for the VLC based fog architecture proposed in [16]. The aim of this study is to utilize the processing nodes adjacent to users, to save network power that otherwise will be consumed by serving high data rate demands in remote conventional cloud data center. We develop a mixed integer linear programming (MILP) model to minimize the power consumption of both processing and networking by optimizing the allocation of processing demand in the fog computing architecture.

The rest of this paper is organized as follows. Section 2 describes the proposed PON based backhaul network architecture and introduces the MILP model to optimize the allocation of processing demands. Section 3 presents the results and discuss them, and Section 4 provides the conclusions.

## 2. PON BACKHAUL NETWORK FOR VLC BASED FOG COMPUTING ARCHITECTURE

In this work, we extend the fog architecture studied in [22] by increasing the number of rooms in a building to four and introducing PON-based backhaul architecture to support the communication within the building. As shown in

Figure 1, each room has eight VLC access points (APs) serving eight users. As in [22], all APs use the red wavelength and each AP provides a data rate of 2.5 Gbps for each user. Note that for the shade of white illumination (in VLC) selected during day time, the red colour dominates which offers a higher transmit power. Therefore, the red colour is considered to connect the user devices (UDs) with the APs. Each AP is attached to an optical network unit (ONU) equipped with tuneable transceivers.

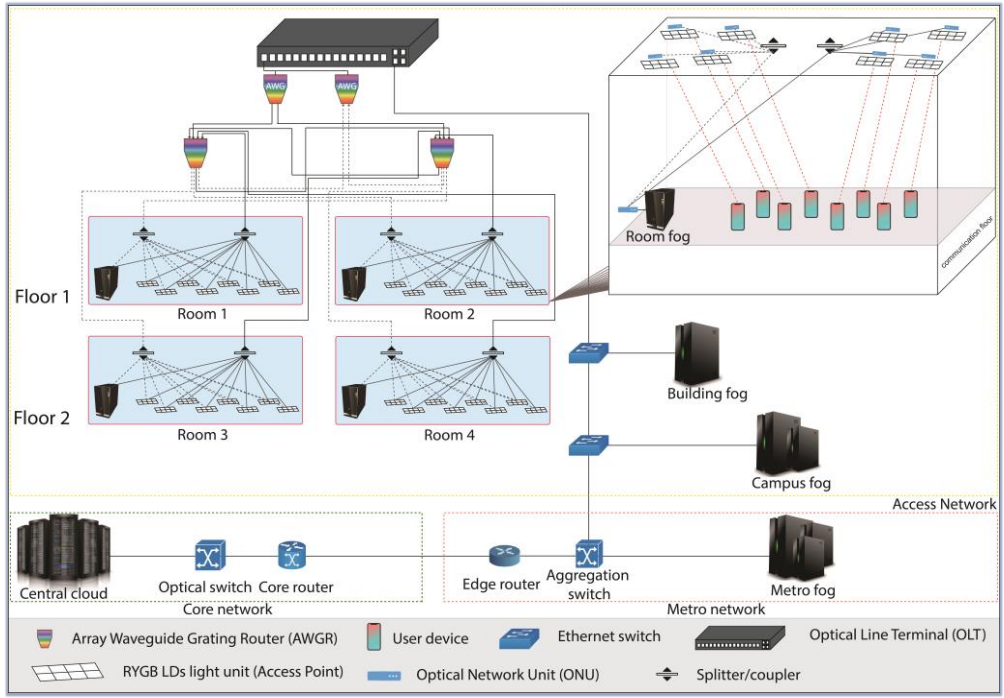

*Figure 1: The proposed PON backhaul network architecture.*

The fog computing resources consists of the idle processing resources in UDs, a single fog server in each room, a building fog node, a campus fog node and fog nodes at the access and metro networks. Processing resources are also available at the cloud data center. In this work, we consider a PON based architecture adopted from the PON design in [13]. A hybrid wavelength division multiplexing (WDM) – time division multiplexing (TDM) PON is deployed where several wavelengths are used to facilitate communication inside the building and each wavelength is shared through TDM among the ONUs in each room. In each room, the ONUs are connected to a splitter and a coupler for upstream and downstream communication, respectively. Two 4×4 arrayed waveguide grating router (AWGRs) are used to provide connectivity between the rooms. The splitters and couplers of each room are connected to an AWGR input port and an AWGR output port, respectively. The AWGRs facilities connectivity between APs within a room or in different rooms. Additionally, The AWGRs are connected to an OLT port to connect the building access network to higher network levels (i.e., metro and core networks). Five wavelengths are used to provide the communication within the access network. A distinct wavelength is used for communication between the APs in the same room. Additionally, a wavelength is used to connect the APs in the rooms to the OLT. The remaining three wavelengths are used to provide connectivity between the rooms.

We developed a MILP model to optimally allocate the processing resources to serve demands with minimum processing and networking power consumption. The developed MILP model is subject to a number of constraints. Theses constraints include a constraint to ensure that at each demand is served by one processing node. However, more than one task can be assigned to the same processing node. Moreover, a constraint ensures that the demands served by a node does not exceed its processing capacity. The model is also subject to the traffic flow conservation constraint, a set of constraints to ensure wavelength continuity in connections between source and destination pairs.

### 3. RESULTS AND DISCUSSION

In this section, we evaluate the performance of the model by examining a scenario with eight users in each room where two UDs are generating demands and the rest of the UDs offer their idle devices to work as processing nodes. Each room has eight VLC APs, each connecting a single user. Each user generates a single task. The demands processing load takes values in the range of 6 - 20 GFLOPs. The traffic demand is related to the processing demand by the Data Rate Ratio (DRR) (the ratio of the traffic demand in Gbps to the processing demand in GFLOPs). In this work, we study applications that require intensive communication and processing such as video gaming applications. To represent these applications, a DRR of 0.05 is considered (i.e., the traffic demands take values in the range 0.3 - 1 Gbps). Table 1 and Table 2 summarize the parameters of the processing nodes and networking devices, respectively.

The performance of the PON backhaul network is compared with a Spine-and-Leaf (SL) backhaul based network, where a leaf switch is used to connect the APs and the room fog server in each room as shown in Figure 2. A total of four leaf switches are connected with two spine switches. The spine switches are then connected to a gateway router to link the access network with the metro network.

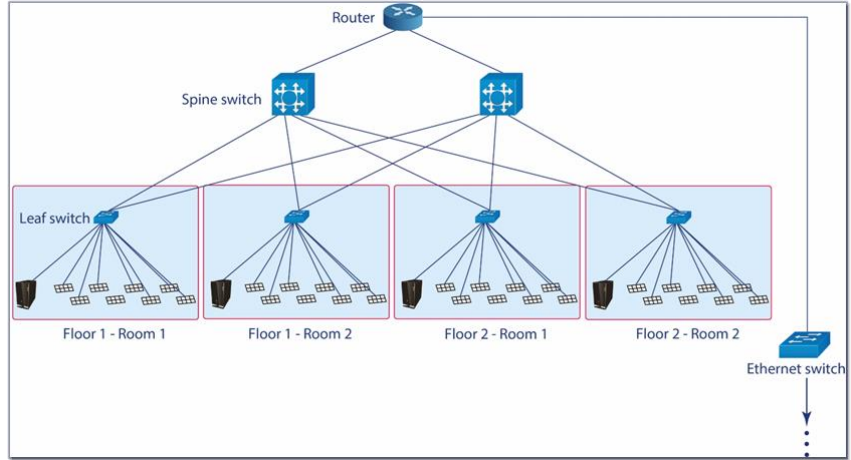

Figure 2: Spine-and-Leaf architecture.

TABLE 1: PROCESSING DEVICES PARAMETERS

| Parameter | Value |
|---|---|
| Processing capacity of the user device (ARM Cortex A53) | 12.888 GFLOPs [24] |
| Processing capacity of the room fog server (Core i3-6006U) | 64 GFLOPs [25] |
| Processing capacity of the building fog server (Intel Xeon Processor E3-1220 v2) | 99 GFLOPs [26] |
| Processing capacity of the campus fog server (Intel Xeon Processor E5-2440 v2) | 121.6 GFLOPs [26] |
| Processing capacity of the metro fog server (Intel Xeon Processor E5-4650 v3) | 403.2 GFLOPs [26] |
| Processing capacity of the cloud server (Intel Xeon Platinum 8280 Processor) | 1612.8 GFLOPs [26] |
| Maximum power consumption of each user device | 18 Watts [27] |
| Maximum power consumption of the Room fog server | 65 Watts [28] |
| Maximum power consumption of the building fog server | 305 Watts [29] |
| Maximum power consumption of the campus fog server, | 350 Watts [30] |
| Maximum power consumption of the metro fog server | 750 Watts [31] |
| Maximum power consumption of the cloud server | 1100 Watts [32] |

TABLE 2: NETWORKING DEVICES PARAMTERS

| Network device | Maximum power consumption (Watts) | Idle power consumption (Watts) | Capacity (Gbps) |
|---|---|---|---|
| Access Point | 7.2 [22] | 4.32 | 2.5 [22] |
| ONU | 15 [33] | 9 | 10 [33] |
| OLT Line card | 300 [34] | 180 | 160 [34] |
| Ethernet switch | 435 [35] | 261 | 240 [35] |
| Aggregation switch | 435 [35] | 261 | 240 [35] |
| Edge router | 750 [36] | 450 | 480 [36] |
| Optical switch | 63.2 [37] | 37.92 | 100 [37] |
| Core router | 344 [38] | 206.4 | 3200 [38] |
| Leaf switch | 508 [39] | 304.8 | 480 [39] |
| Spine switch | 660 [40] | 360 | 1440 [40] |
| router | 344 [38] | 206.4 | 3200 [38] |

Furthermore, serving processing demand in the proposed fog computing architecture is compared to the case when the central cloud serves all the demands. Figure 3 shows the processing workload allocation to the UDs, fog servers in rooms 1, room 2, room 3, and room 4 (i.e., r1RF, r2RF, r3RF, and r4RF, respectively), the building fog (BF), the campus fog (CF), the metro fog (MF) and cloud resources (CC) in both PON-based architecture and the SL architecture. As observed in Figure 3 for both architectures, all the demands are exclusively served within the rooms, without the need to activate further remote fog units in the access or metro networks. At 6 GFLOPs in the PON-based architecture, the demands are served only in the fourth room fog server (r4RF). Accessing the room fogs and the UDs results in a similar network power consumption due to the nature of the passive network. However, consolidating demands into a single fog room server is more efficient than activating multiple UDs as it results in using less total idle power. In other words, when a single room fog serves all demands, the amount of idle power consumed is reduced. It is worth noting that the selection of which room fog to activate is random and will result in the same total power consumption due to the use of homogeneous specifications for the room fogs.

In contrast, the SL backhaul network shows a different trend in allocating processing resources at 6 GFLOPs. While room fog servers are more efficient in terms of processing, their idle power consumption is significantly higher compared to UDs. Because of the high networking power consumption required to access a single room fog, the optimal allocation decision involves activating UDs in all rooms instead of relying on a single room fog. Therefore, one UD is activated in each room to host two tasks instead of activating four room fogs. At 7 and 8 GFLOPs, the allocation in the PON-based architecture remains the same (i.e., demands allocated to one room fog server). However, for SL at 7 GFLOPs, all room fog servers (r1RF, r2RF, r3RF, r4RF) are activated. Although consolidating the demands in one fog room server can save processing power, it will lead to a significant increase in the network power consumption due to the need to pass through a second level of spine switches. Additionally, the UDs are avoided because of the need to activate more than one UD in each room. To justify the decision of the model, it is important to highlight that the processing capacity of each UD is 12.88 GFLOPs. Since each device will only be able to accommodate one task, instead of activating two UDs in each room, it is more efficient to activate the room fog server in each room. The allocation in the SL architecture continues to follow the same trend (i.e., the demands are served from all room fog servers) until 20 GFLOPs. In the PON-based architecture at 9 GFLOPs, the room fog server capacity is exhausted, and hence, one idle user UD is activated to serve the remaining demands in addition to the room fog server. Note that at 10 GFLOPs, another room fog server is utilized to serve the demands (r4RF) instead of activating another two UDs, and this trend remains the same for the demands up to 15 GFLOPs. At 16 GFLOPs, a third room fog server is activated, and three fog room servers are sufficient to serve demands up to 20 GFLOPs.

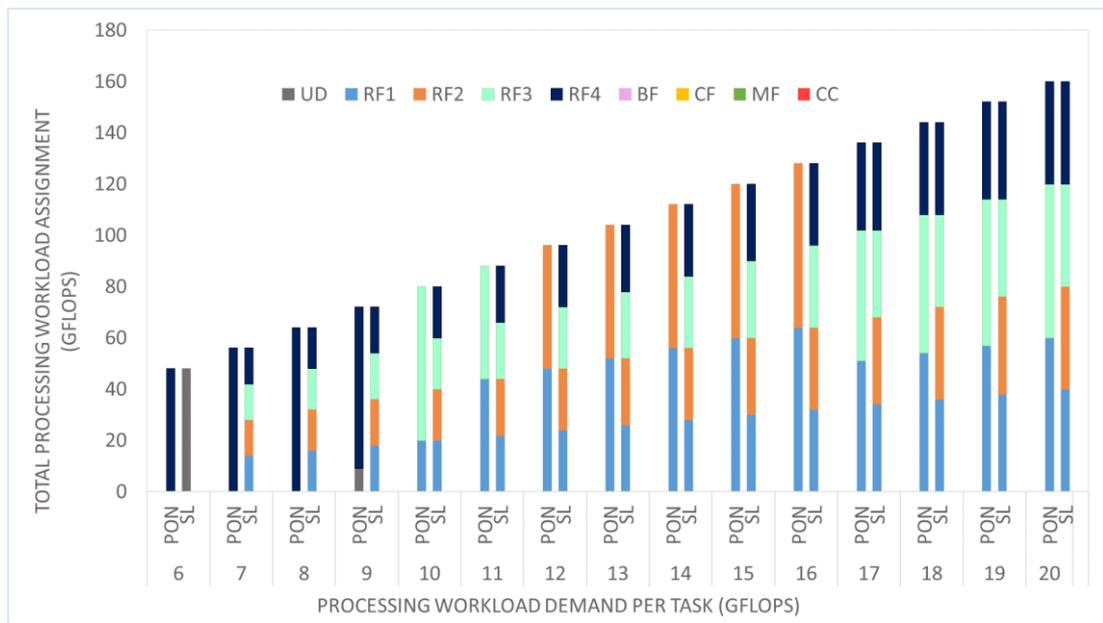

*Figure 3: Processing workload allocation in PON-based architecture and SL architecture for the considered scenario.*

Figure 4 shows a comparison of the power consumption for optimized workload allocation between the PON-based architecture, SL architecture and serving the demands from the cloud only. Figure 4.a presents a comparison of the processing power consumption. The obtained results show that the PON-based architecture is notably more efficient in terms of processing compared to the SL architecture as a result of consolidating the workloads into fewer processing nodes. The processing savings achieved are up to 38%. The processing in the cloud is extremely efficient, nevertheless, it increases the networking power consumption by 92% compared to serving the demand locally with the support of the PON architecture as can be observed in Figure 4.b. Compared to the SL based architecture, the PON-based architecture saves 90% of the networking power consumption. Figure 4.c shows the total power consumption including both processing and networking. Total savings of up to 86% and 84% can be achieved when deploying PON-based architecture compared to serving the demands from the cloud and using SL, respectively.

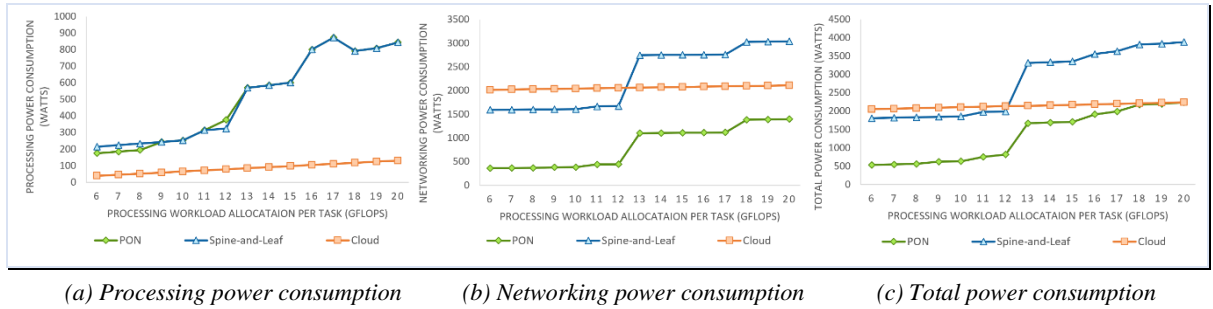

*(a) Processing power consumption     (b) Networking power consumption     (c) Total power consumption*

*Figure 4: Power consumption comparison between PON-based architecture, SL architecture, and serving the demands from the cloud for the considered scenario.*

## 4. CONCLUSIONS

In this paper, we proposed an energy efficient PON backhaul network for a VLC based fog architecture where fog resources within a building complement fog nodes at the access and metro networks and the central cloud data center. We studied the allocation of processing and data rate intensive applications in the proposed architecture. We developed a MILP model to minimize the total power consumption of serving demands by optimizing of allocation of resources to demands. The resource allocation results show the ability of the PON backhaul network to consolidate demands in fewer nodes as a resulting of its ability to connect users and room fogs efficiently compared to an architecture based on a SL based backhaul. Total savings of up to 84% can be achieved by deploying the proposed PON architecture compared to SL architecture. Additionally, relative to allocating the demands in the cloud, the optimal allocation of the proposed architecture can save 86% of the total power consumption.

**ACKNOWLEDGEMENTS**

The authors would like to acknowledge funding from the Engineering and Physical Sciences Research Council (EPSRC) INTERNET (EP/H040536/1), STAR (EP/K016873/1) and TOWS (EP/S016570/1) projects. For the purpose of open access, the authors have applied a Creative Commons Attribution (CC BY) licence to any Author Accepted Manuscript version arising. All data are provided in full in the results section of this paper.